\documentclass[3p]{elsarticle}

\usepackage[colorlinks,hyperindex]{hyperref}
\usepackage[utf8x]{inputenc}
\usepackage{float}
\usepackage{latexsym}
\usepackage{graphicx}
\usepackage{amssymb}
\usepackage{amsmath}
\usepackage{braket}
\journal{Optik}

\begin{document}

\begin{frontmatter}

\title{Generation of matter threefold star state by dynamical transfer from photonic counterpart}
\author{J. P. Restrepo Cuartas\corref{cor1}}
\ead{jprestrepocu@unal.edu.co}
\author{M. Cardona-Rinc\'on}
\author{H. Vinck-Posada}
\address{Departamento de F\'isica, Universidad Nacional de Colombia, 111321, Bogot\'a, Colombia}
%\address[label2]{Programa de F\'isica, Universidad del Quind\'io, 630004, Armenia, Colombia\fnref{label4}}
\cortext[cor1]{Corresponding author}

%\ead[url]{author-one-homepage.com}
%
\begin{abstract}
In this paper, the dynamics of a system composed by an assembly of two-level atoms coupled to a single-mode cavity is theoretically studied considering the Tavis-Cummings  Hamiltonian. Motivated by recent experimental results in non-linear phenomena on squeezed-like states, we have found that a threefold star state for the whole set of emitters can be dynamically generated from the analogue photonic state; beyond this, we numerically demonstrate, that this transfer of star-like states critically depends with the number of emitters. Quantum statistical properties of the system are obtained by computing the second-order correlation function, the linear entropy and the Wigner function for the reduced density matrices between radiation and collective matter components.
\end{abstract}


\end{frontmatter}

\section{Introduction}\label{intro}

The collective emission of light by an assembly of
atoms has been addressed in early quantum optics 
days by Dicke~\cite{dicke54a}. This subject was then
carried forward through work on the exact solution for
an N-molecule--Radiation system, i.e., the so-called
Tavis-Cummings (TC) model~\cite{tavis68a}. This topic
was then developed by the outstanding work of 
Hepp and Lieb where they introduced the acclaimed
Dicke-model second-order phase 
transition~\cite{Hepp73a}---a phase transition for
the N-emitters coupled with photons in thermal
equilibrium that goes from normal to superradiance.
The theme rapidly becomes a successful area of 
research~\cite{Hepp73b,Wang73,Carmichael73}.
Nowadays, multiple studies that involve the
dynamical description for the Dicke and 
Tavis-Cummings models have been proposed. The 
challenges in this area have been to identify the
dynamical quantum behaviour close to classical
limit~\cite{Lerma-Hernandez2018,Tsyplyatyev10},
to characterise the superradiant to subradiant
phase transition by dark state cascades~\cite{Gegg18},
to study the effects of the dynamical properties 
related with the finite size in the Dicke model
while is coupled to a thermal
reservoir~\cite{Imai19}, to determine the
dynamical entanglement properties~\cite{Mao16,He15},
and the quantum state preparation~\cite{Joshi16}. 
In the past, it has been explored how to describe
the ground state to get the main features of the
BCS-BEC phase transition~\cite{eastham01a,Yamaguchi13,Horikiri16, byrnes10a,Chen05}; to do
that, the authors used a variational approach 
taking into account a pure state built up by the
Kronecker product of a BCS-like state and a light
coherent state.  

We focus in this paper on the Hamiltonian dynamics of
threefold star light states which does not leads to
dissipation and decoherence phenomena~\cite{carmichael93a,Breuer02a}. The transfer of this kind of states onto an assembly of emitters~\cite{Scully07a,Sutherland16a} is analysed. The Schrödinger equation rules the dynamical behaviour for many 
isolated systems in quantum optics~\cite{alicki89a,briegel93a,perea04a},
quantum  measurement theory~\cite{schlosshauer04a},
quantum statistical  mechanics, quantum information
science~\cite{imamoglu99a,you05a}, along with others. 

The goals of most recent research are  motivated by
the need for a systematic theoretical approach 
that accounts for the quantum dynamical features and
light-matter interaction properties of the Dicke 
superradiance~\cite{dicke54a,gross82a}---particularly, the single  photon  superradiance---~\cite{Scully07a,Sutherland16a,Wu17}. 

The paper is organised as follows: in section 2, we describe
the physical system and introduce the theoretical framework which includes the main features of the characterisation tools like the Wigner function of matter
onto the Bloch sphere. Then, in section 3, we analysed the dynamics
of a threefold star state of light and its transfer to the matter 
component of the system and its dependence on the number of emitters. Finally, in section 4, we
summarise and conclude.

\section{Theoretical model}\label{sec:formalism}
\begin{figure}[ht]
\centering
\includegraphics[width=0.97\textwidth]{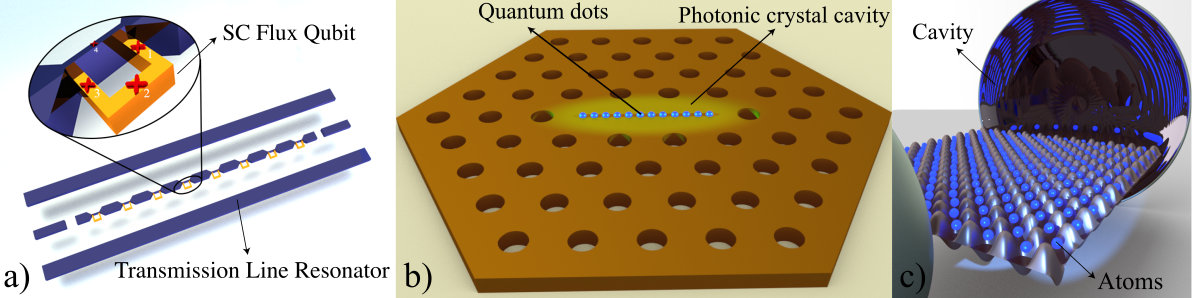}
\caption{Panel a) Sketch of the circuit QED realisation 
of Dicke model. Linear coupling is assumed between the 
cavity and each two-level emitter~\cite{Feng15a}. b)  Schematic rendering of a photonic crystal with $\mathrm{N}$ emitters
embedded~\cite{Manzoni17}. c) Experimental scheme for preparing ultracold two-level atoms inside high- optical Fabry-Perot resonators~\cite{Ritsch13}.}
\label{system}
\end{figure}

\noindent
The theoretical model involves an assembly
of quantum emitters (QEs) coupled to a single cavity mode
through the TC Hamiltonian that was successful for describing multiple experimental realisations as it is shown schematically in Fig. \ref{system}.
The TC Hamiltonian which describes the system is
composed by a cavity mode with frequency $\omega_c$,
the N two-level atoms with frequencies
$\omega_{a}=\omega_c-\Delta$, and the 
light--matter coupling due to a linear dipolar
interaction, can be written as
\begin{equation}
      H_\mathrm{TC}= \omega_{\mathrm{c}} a^\dag a + \omega_\mathrm{a}{J_\mathrm{z}}+\frac{\lambda}{\sqrt{\mathrm{N}}}(a^\dag J_ - + a J_+
      ),
      \label{eq:HTC}
\end{equation}
where $ a^\dag$ ($ a$) is the boson creation
(annihilation) operator in the Fock basis, where $\hbar=1$. $ J_z,
J_{\pm}$  are collective angular momentum operators for the set of N two-level emitters, and
$\lambda$ is the matter--field strength 
coupling \cite{garraway11}.
%%%
%%%
%
\subsection{Angular momentum states}%
To describe the dynamics of the matter component for the system, it is necessary to define the set of two-level operators ($\sigma^{\mathrm{i}}_\mathrm{z}$, $\sigma^{\mathrm{i}}_{\pm}$), which can be written in terms of the matter bare basis ($\ket{\mathrm{g}_{i}}, \ket{\mathrm{e}_{i}}$), as
\begin{align}
       \sigma_\mathrm{z}^\mathrm{i}&=\ket{\mathrm{e}_\mathrm{i}}\bra{\mathrm{e}_\mathrm{i}}-\ket{\mathrm{g}_\mathrm{i}}\bra{\mathrm{g}_\mathrm{i}},\,\,\sigma_+^\mathrm{i}=\ket{\mathrm{e}_\mathrm{i}}\bra{\mathrm{g}_\mathrm{i}},\,\, \text{and} \,\, \sigma_-^\mathrm{i}=\ket{\mathrm{g}_\mathrm{i}}\bra{\mathrm{e}_\mathrm{i}},
    \label{eq:SigmaOps}
\end{align}
where the exterior product $\ket{\Psi}=\bigotimes^{\mathrm{N}}_{\mathrm{i}}\ket{\alpha_{\mathrm{i}}}$ form the $\mathrm{N}$-emitters Hilbert space, with $\alpha_{\mathrm{i}}=\mathrm{g}_{\mathrm{i}},\mathrm{e}_{\mathrm{i}}$, and the operators must be understood as $\sigma^{\mathrm{i}}_{\mathrm{\mu}}:=\dotsm \otimes\mathbb{I}^{\mathrm{i}-1}\otimes\sigma^{\mathrm{i}}_{\mathrm{\mu}}\otimes\mathbb{I}^{\mathrm{i}+1}\otimes\dotsm$, and $\mu=\mathrm{z},\, \pm$.
The collective angular momentum operators of the TC Hamiltonian in Eq. (\ref{eq:HTC}) can be written in terms of the $\mathrm{N}$-emitters operators as follows
\begin{equation}
       J_{\mathrm{z}}=\frac{1}{2}\sum_{\mathrm{i}=1}^{\mathrm{N}}{\sigma_\mathrm{z}^\mathrm{i}}\,\,\,\, \text{and} \,\,\,\,  J_{\mathrm{\pm}}=\frac{1}{2}\sum_{\mathrm{i}=1}^{\mathrm{N}}{\sigma_\mathrm{\pm}^\mathrm{i}}.
      \label{eq:JOps}
\end{equation}
The set of angular momentum operators 
satisfy the commutation relations
$[ J_+, J_-]=2 J_\mathrm{z}$. The eigenstates of $J^{2},\,J_{\mathrm{z}}$ as it is well-known in the literature, the
Dicke basis \cite{dicke54a,Arecchi72,Yamamoto99}, can be generated from the vacuum state  ($J_-\ket{-\mathrm{J}}=0$),
that takes the form
\begin{equation}
      \ket{\mathrm{M}}:=\ket{\mathrm{M},\mathrm{J}}=\frac{1}{(\mathrm{M}+\mathrm{J})!}\binom{2\mathrm{J}}{ \mathrm{M}+\mathrm{J}}^{-\frac{1}{2}}J_+^{\mathrm{M}+\mathrm{J}}\ket{-\mathrm{J}},
      \label{DickeState1}
\end{equation}
bear in mind that the eigenvalues of $J^{2}$ and $J_{\mathrm{z}}$ are $\mathrm{J}(\mathrm{J}+1)$ and $\mathrm{M}$, respectively, with
$\mathrm{M}=-\mathrm{J},-\mathrm{J}+1,\dots,\mathrm{J}$. Moreover, the non-Hermitian operator $J_{+}$ ($J_{-}$) raises (decreases) the eigenvalue $\mathrm{M}$ to $\mathrm{M}+1$ ($\mathrm{M}-1$). The raising/lowering operators satisfy the relation $J_{\pm}\ket{\mathrm{M}}=\sqrt{\mathrm{J(\mathrm{J}+1)}-\mathrm{M}(\mathrm{M}\pm 1)}\ket{\mathrm{M}\pm 1}$.
\subsection{Equations of Motion}
%%%
The dynamics of the system is ruled by the Schr\"odinger equation $i\partial_{\mathrm{t}}\ket{\psi}=H_{\mathrm{TC}}\ket{\psi}$. We build the Hilbert space starting from the uncoupled eigenstates of the system defined as $\ket{\mathrm{M},\mathrm{n}}:=\ket{\mathrm{M}}\otimes\ket{\mathrm{n}}$, where $\ket{\mathrm{n}}$ is a Fock state and $\ket{\mathrm{M}}$ is a state vector in the Dicke basis. We can span the global state of the system as a linear combination of the latter states as $\ket{\psi}=\sum_{\mathrm{M=-J}}^{\mathrm{J}}\sum_{\mathrm{n=0}}^{\infty}{\mathrm{C_{\mathrm{M}\mathrm{n}}}\ket{\mathrm{M},\mathrm{n}}}$. Replacing this state into the Schr\"odinger equation, and projecting onto the global basis, it can be obtained the relation
\begin{align}
\dot{\mathrm{C}}_{\mathrm{Mn}}(\mathrm{t})=&-i\left(\omega_\mathrm{c}\mathrm{n}+\omega_\mathrm{a}\mathrm{M}\right)\mathrm{C}_{\mathrm{Mn}}(\mathrm{t})\\ \nonumber
    &-i\frac{\lambda}{\sqrt{\mathrm{N}}}\sqrt{\mathrm{n}+1}\sqrt{\mathrm{J(J+1)}-\mathrm{M(M-1)}}\mathrm{C}_{\mathrm{M-1n+1}}(\mathrm{t})\\ \nonumber
    &-i\frac{\lambda}{\sqrt{\mathrm{N}}}\sqrt{\mathrm{n}}\sqrt{\mathrm{J(J+1)}-\mathrm{M(M+1)}}\mathrm{C}_{\mathrm{M+1n-1}}(\mathrm{t}).
\end{align}    
This equation is solved by truncating the photonic basis until a maximal photon number. In particular, we are interested in the following initial condition
\begin{equation}
    \ket{\psi(\mathrm{t}=0)}=\eta\ket{\mathrm{M},0} + \beta \ket{\mathrm{M},3},
\end{equation}
here, $|\eta|^{2}+|\beta|^{2}=1$. The photonic state
$\ket{\bigstar}=\eta\ket{0}+\beta\ket{3}$ is known in the
literature like the threefold star state \cite{JCLopez14a,gevorgyan13a,antonosyan11a}. The
preparation of such a state inside a cavity can be made by
a photonic crystal capable of enhancing the harmonic
generation produced by either a $\chi^{(2)}$ or $\chi^{(3)}$ nonlinearity. In the case of the former, the threefold star state results as a consequence of the coupling between an emitter embedded into the cavity with a $\omega_0$ mode of the electromagnetic field, and two SPDC processes which produces  two additional cavity modes with energies: $\omega_0 \rightarrow \omega_1 + \omega_2$ and $\omega_2 \rightarrow \omega_1 + \omega_1$.
We can build up the state operator from the state vector as $\rho=\ket{\psi}\bra{\psi}$. The matrix representation of the state operator (density matrix) can be obtained using the uncoupled eigenstates basis span coefficients  $\rho_{\mathrm{Mn,Ls}}=\mathrm{C_{\mathrm{M}\mathrm{n}}}\mathrm{C_{\mathrm{L}\mathrm{S}}^\ast}$. Finally, we calculate the reduced density operators  $\rho_{\mathrm{ph}}=\mathrm{Tr}_\mathrm{D}(\rho)$ and $\rho_{\mathrm{D}}=\mathrm{Tr}_\mathrm{ph}(\rho)$ for the photonic and matter components, respectively.

\subsubsection{Hamiltonian eigenstates} 

The dynamics of the system when a star state of light is prepared as initial condition ---along with all emitters in their ground state ($\ket{-\mathrm{J}}$)--- is restricted to the states $\ket{-\mathrm{J},3}$, $\ket{-\mathrm{J}+1,2}$, $\ket{-\mathrm{J}+2,1}$, $\ket{-\mathrm{J}+3,0}$. Obviously, the global ground state $\ket{-\mathrm{J},0}$ is taken into account in the definition of the star state, but it remains dynamically uncoupled. Hamiltonian in this basis is written as

 \begin{equation}\label{eq:hamatrix}
     \left(
\begin{array}{cccc}
 3\Delta-\Delta_0\mathrm{J} &   \mathrm{g} \sqrt{6\mathrm{J}} & 0 & 0 \\
  \mathrm{g} \sqrt{6\mathrm{J}} & 2\Delta -\Delta_0(\mathrm{J}-1) & 2  \mathrm{g}
   \sqrt{2 \mathrm{J}-1} & 0 \\
 0 & 2  \mathrm{g} \sqrt{2\mathrm{J}-1} & \Delta -\Delta_0(\mathrm{J}-2) &  \mathrm{g} \sqrt{6(\mathrm{J}-1)}
   \\
 0 & 0 &  \mathrm{g} \sqrt{6(\mathrm{J}-1)} & -\Delta _0(\mathrm{J}-3) \\
\end{array}
\right).
 \end{equation}

This matrix is a function of the number of emitters in the assembly through the angular momentum eigenvalue $\mathrm{J}$. The limit for a large number of emitters can be achieved by making $\mathrm{J}\rightarrow\infty$. Therefore the Hamiltonian matrix takes the following form

 \begin{equation}\label{eq:hamatrixlim}
     \left(
\begin{array}{cccc}
 -\Delta_0\mathrm{J} &  \mathrm{g} \sqrt{6\mathrm{J}} & 0 & 0 \\
  \mathrm{g} \sqrt{6\mathrm{J}} & -\Delta_0\mathrm{J} & 2  \mathrm{g}
   \sqrt{2 \mathrm{J}} & 0 \\
 0 & 2  \mathrm{g} \sqrt{2\mathrm{J}} & -\Delta_0\mathrm{J} &  \mathrm{g} \sqrt{6\mathrm{J}}
   \\
 0 & 0 &  \mathrm{g} \sqrt{6\mathrm{J}} & -\Delta _0\mathrm{J} \\
\end{array}
\right).
 \end{equation}
As the manifold (defined as $N_{\mathrm{exc}}=\mathrm{J}\mathbb{I}+J_{\mathrm{z}}+a^{\dagger}a$) is conserved we can analyse the general dynamics restricted to each sub-block of the complete Hamiltonian, even for a large number of emitters. To study eigenvalues and eigenvectors by blocks is equivalent to study the general eigenvalue problem of the Hamiltonian. 
\subsubsection{Quasi-probability representation of two-level assembly states and photons, second-order correlation function, and linear Entropy}
\label{sec:WignerFunction}

The representations on the Bloch sphere of the density operator are 
extremely useful to exhibit the main features of quantum correlations~\cite{pezze18a}.
We consider only symmetric states of $\mathrm{N}$ two-level emitters.
There are a plethora of pictures in the form of quasi-probability densities which are based on the decomposition of a general density operator  $\rho_{\mathrm{D}}=\sum_{\mathrm{k}=0}^{2\mathrm{J}} \sum_{\mathrm{q=-k}}^\mathrm{k} \sigma_{\mathrm{k q}}T_{\mathrm{k q}}^{(\mathrm{J})}$  into the ortho-normalised spherical tensor (or multipole) operators $T_{\mathrm{k q}}^{(\mathrm{J})}=\sum_{\mu,\mu'=-\mathrm{J}}^\mathrm{J} (-1)^{\mathrm{J}-\mu'} \braket{\mathrm{J},\mu;\mathrm{J},-\mu'|\mathrm{k,q}} \ket{\mathrm{J},\mu_\mathrm{z}}\bra{\mathrm{J},\mu'_\mathrm{z}}$
defined in terms of Clebsch-Gordan coefficients, with  $\sigma_{\mathrm{k q}}=\mathrm{Tr}[\rho_{\mathrm{D}}(T_{\mathrm{k q}}^{(\mathrm{J})})^\dagger ]$. Therefore, this fact allows to define the family of spherical functions 
\begin{equation} \label{eq:sphfunc}
	\mathrm{f}(\vartheta,\varphi) =
	\sqrt{\frac{\mathrm{N} + 1}{4 \pi}}
	\sum_{\mathrm{k}=0}^{\mathrm{N}} \mathrm{f}_\mathrm{k} \sum_{\mathrm{q=-k}}^\mathrm{k} \sigma_{\mathrm{k q}} \mathrm{Y}_{\mathrm{k q}}(\vartheta,\varphi),
\end{equation}
in terms of spherical harmonics $\mathrm{Y}_{\mathrm{k q}}(\vartheta,\varphi)$. The Wigner quasi-probability distribution $\mathrm{W}(\vartheta,\varphi)$ \cite{Wigner32} is obtained by the case $\mathrm{f}_\mathrm{k}=1$ in Eq.~\eqref{eq:sphfunc}.  It is well known that for all purposes the Wigner quasi-probability distribution is equivalent to the density operator; even though $\mathrm{W}(\vartheta,\varphi)$ is not a true probability density because it can take negative values; this fact is interpreted as evidence of non-classical behaviour. Indeed, the marginals in the phase space are true probability functions. Otherwise, the photonic Wigner function $\mathrm{W}(\alpha)$ at a given phase space point $\alpha$ is defined as the expectation value of the displaced photon number parity operator~\cite{Lutterbach97} $W(\alpha) = 2 \mathrm{Tr}[ D(\alpha)\rho_{\mathrm{ph}} 
  D^{\dagger}(\alpha)\exp(i\pi a^\dagger a)]$, where $D(\alpha) = \exp(\alpha \hat{a}^\dagger - \alpha^{\ast}
\hat{a})$ denotes the displacement operator.  The second-order correlation function is defined in terms of the Hanbury-Brown Twiss (HBT)\cite{hanburybrown56a,hanburybrown57a,hanburybrown58b} experiment, 
which measures using coincidence correlations and anti-correlations by two detectors from a beam of photons prepared in a certain quantum state~\cite{Yamamoto99,Gerry04}. This function is defined as
\begin{equation}\label{eq:secondorder} 
\mathrm{g}^{(2)}(\tau)=\frac{\langle\hat a^\dagger(0)\hat a^\dagger(\tau)\hat a(\tau)\hat a(0)\rangle}{\langle\hat a^\dagger(0)\hat a(0)\rangle^2},
\end{equation}
the main signature of a quantum emission is a vanishing
second-order correlation function $\mathrm{g}^{(2)}(\tau)$ at zero delay $(\tau=0)$. The entanglement measure for a bipartite system can be calculated
in the reduced density operator $\rho_{\mathrm{ph}}=\mathrm{Tr}_\mathrm{D}(\rho)$ or $\rho_{\mathrm{D}}=\mathrm{Tr}_\mathrm{ph}(\rho)$. This means that we must average over all relevant coordinates of the remained subsystem by taking the partial trace. We can use the linear entropy $\mathrm{S}_\mathrm{L}=1-\mathrm{Tr}(\rho_{ph}^2)$, which is also a measure of the purity of the reduced system, to measure the amount of entanglement between the cavity and the emitters. The linear entropy of the reduced state is closely related to the Schmidt decomposition~\cite{Nielsen11}.

\section{Results}\label{Results}
\begin{figure}[H]

\begin{minipage}[c]{\textwidth}
\centering
\hspace{-16pt}\includegraphics[width=0.4\textwidth]{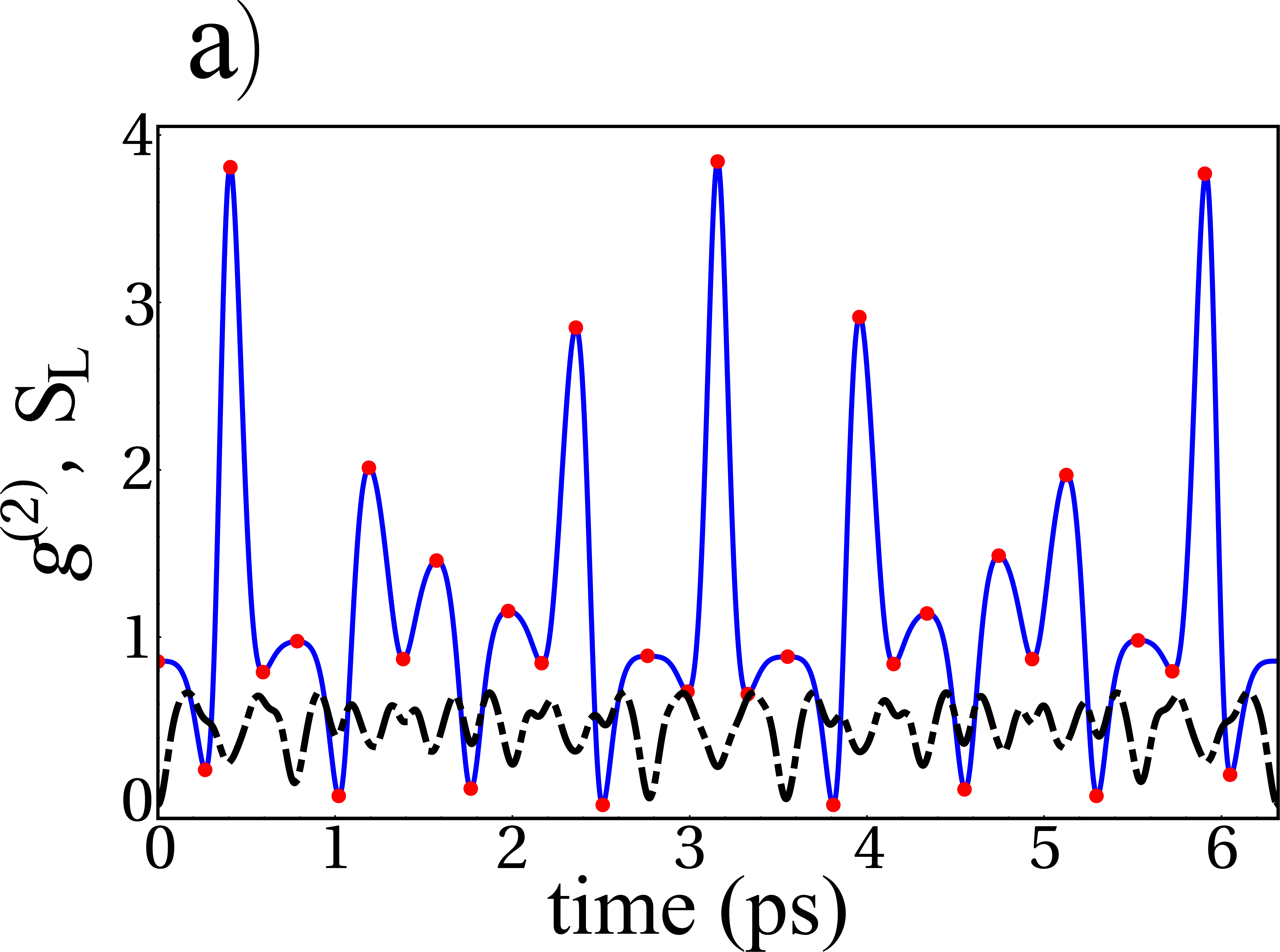}
\vspace{14pt}
\end{minipage}\\
\begin{minipage}[c]{\textwidth}
\centering
\includegraphics[width=0.54\textwidth]{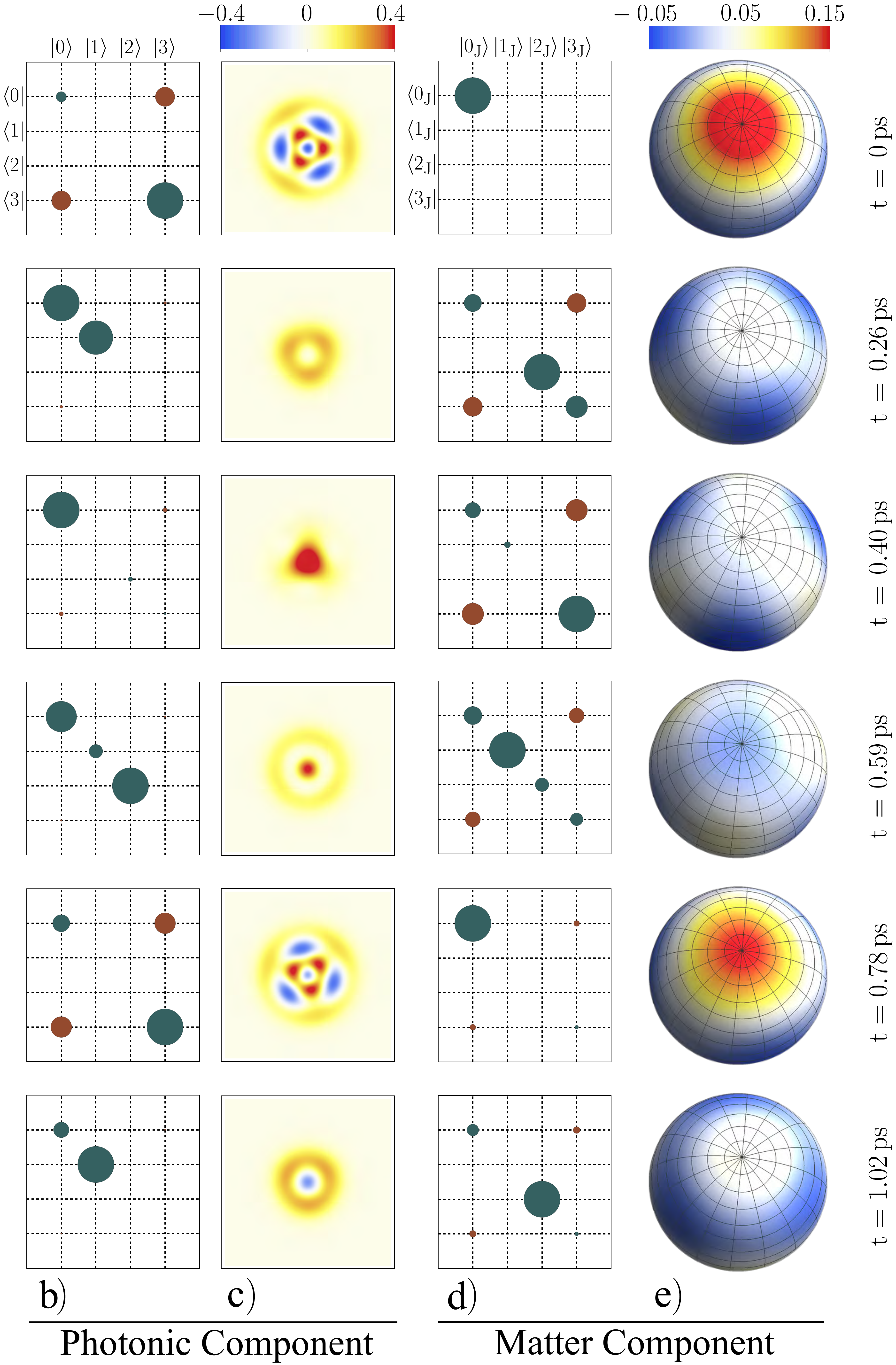}
\end{minipage}
\caption{System dynamics for $\mathrm{N}=3$ emitters and initial condition $\ket{\psi (0)}=\ket{\mathrm{M}}\otimes\ket{\bigstar}$, with $\mathrm{M}=-\mathrm{J}=-3/2$, $\eta=\sqrt{2}/3$ and $\beta=\sqrt{7}/3$. a) evolution of the photonic second-order correlation function $\mathrm{g}^{(2)}$ (blue solid line) and linear entropy $\mathrm{S}_{\mathrm{L}}$ (black dot-dashed line). Red points depict the $\mathrm{g}^{(2)}$ critical values. b) matrix elements for the photonic density operator $\rho_{\mathrm{ph}}$. c) photonic Wigner functions, with $\mathrm{Re}(\epsilon)$ in the abscissa and $\mathrm{Im}(\epsilon)$ in the ordinate, where $\epsilon$ is a point in the phase space. Both abscissa and ordinate axes take values from $-3$ to $3$. d) matrix elements for the matter density operator $\rho_{\mathrm{D}}$. Here, was considered the notation $\ket{\mathrm{i}_{\mathrm{J}}}=\ket{\mathrm{-J+i}}$, with $\mathrm{i}=0,1,2,3$, for the matter states. e) matter Wigner distribution onto the block sphere. In panels b)-e) were considered the times of the first six critical values in the function $\mathrm{g}^{(2)}$.}
\label{graph1}
\end{figure}

\begin{figure}[H]
\begin{minipage}[c]{\textwidth}
\centering
\hspace{-16pt}\includegraphics[width=0.4\textwidth]{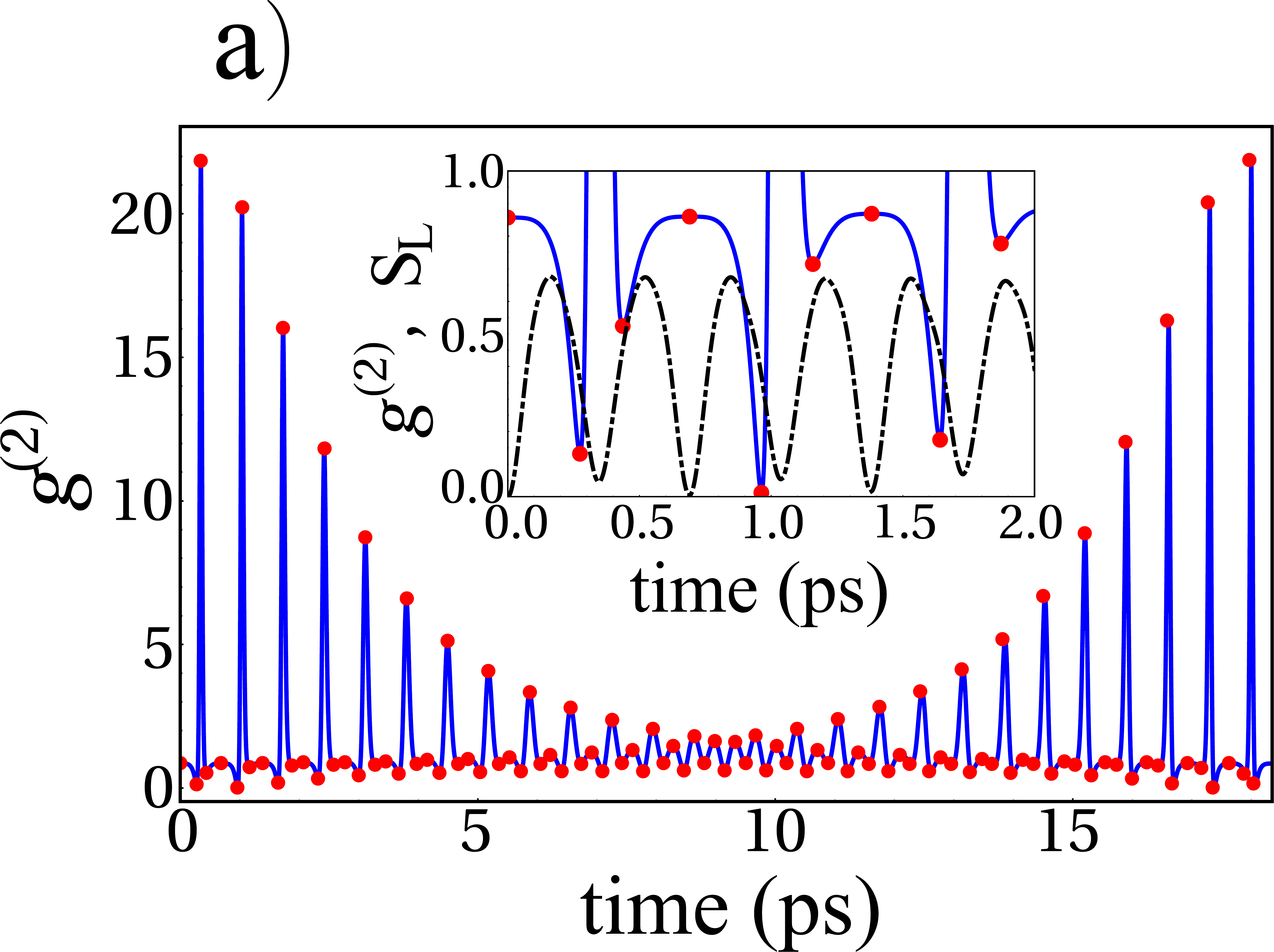}
\vspace{13pt}
\end{minipage}\\
\begin{minipage}[c]{\textwidth}
\centering
\includegraphics[width=0.54\textwidth]{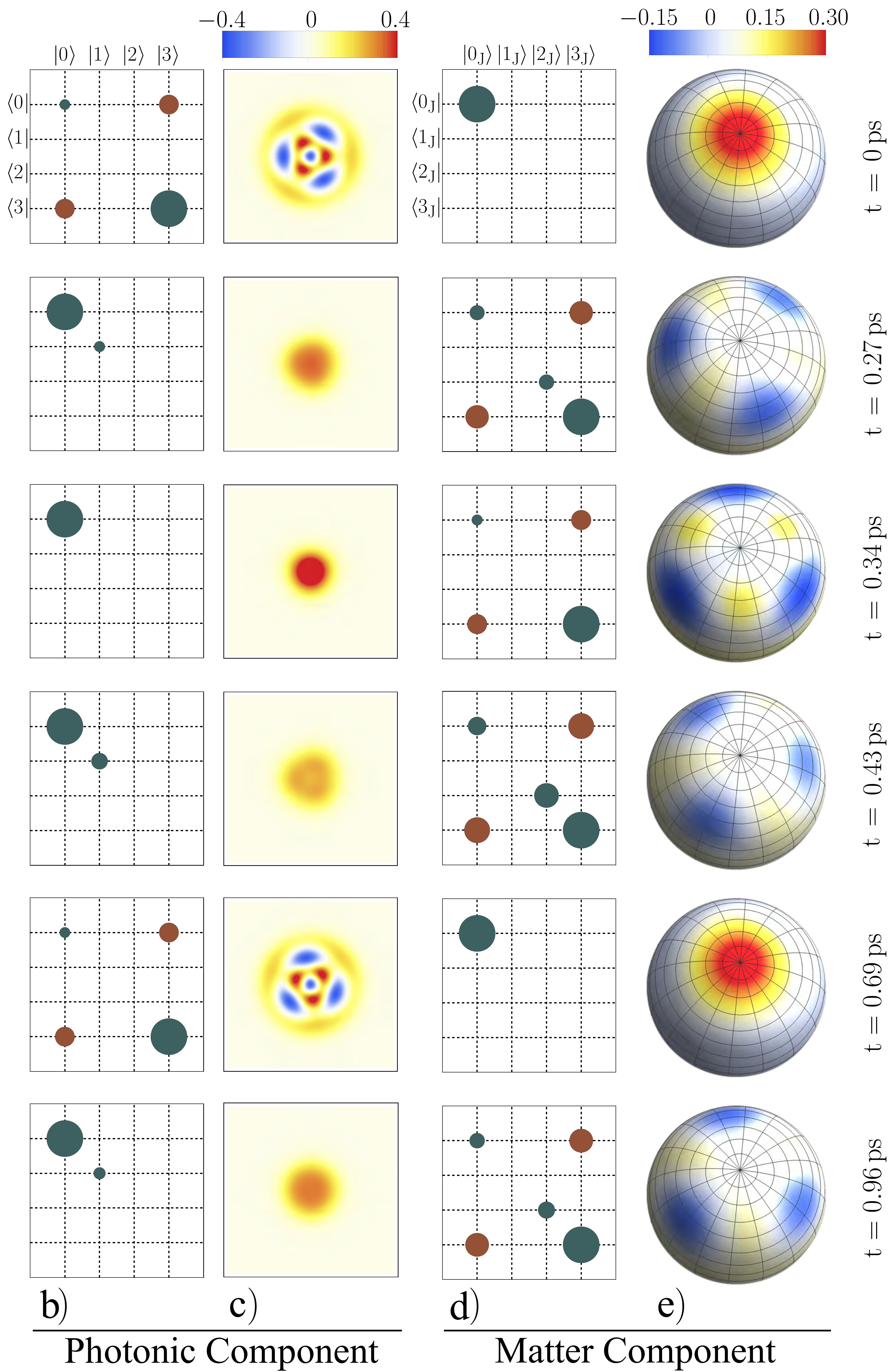}
\end{minipage}
\caption{System dynamics for $\mathrm{N}=6$ emitters and initial condition $\ket{\psi (0)}=\ket{\mathrm{M}}\otimes\ket{\bigstar}$, with $\mathrm{M}=-\mathrm{J}=-3$, $\eta=\sqrt{2}/3$ and $\beta=\sqrt{7}/3$. a) evolution of the photonic second-order correlation function $\mathrm{g}^{(2)}$ (blue solid line). Furthermore, the inset shows the linear entropy $\mathrm{S}_{\mathrm{L}}$ (black dot-dashed line) up to $\mathrm{t}=2$ ps. Red points depict the $\mathrm{g}^{(2)}$ critical values. b) matrix elements for the photonic density operator $\rho_{\mathrm{ph}}$. c) photonic Wigner functions, with $\mathrm{Re}(\epsilon)$ in the abscissa and $\mathrm{Im}(\epsilon)$ in the ordinate, where $\epsilon$ is a point in the phase space. Both abscissa and ordinate axes take values from $-3$ to $3$. d) matrix elements for the matter density operator $\rho_{\mathrm{D}}$. Here, was considered the notation $\ket{\mathrm{i}_{\mathrm{J}}}=\ket{\mathrm{-J+i}}$, with $\mathrm{i}=0,1,2,3$, for the matter states. e) matter Wigner distribution onto the block sphere. In panels b)-e) were considered the times of the first six critical values in the function $\mathrm{g}^{(2)}$.}
\label{graph2}
\end{figure}

\begin{figure}[H]
\begin{minipage}[c]{\textwidth}
\centering
\hspace{-16pt}\includegraphics[width=0.4\textwidth]{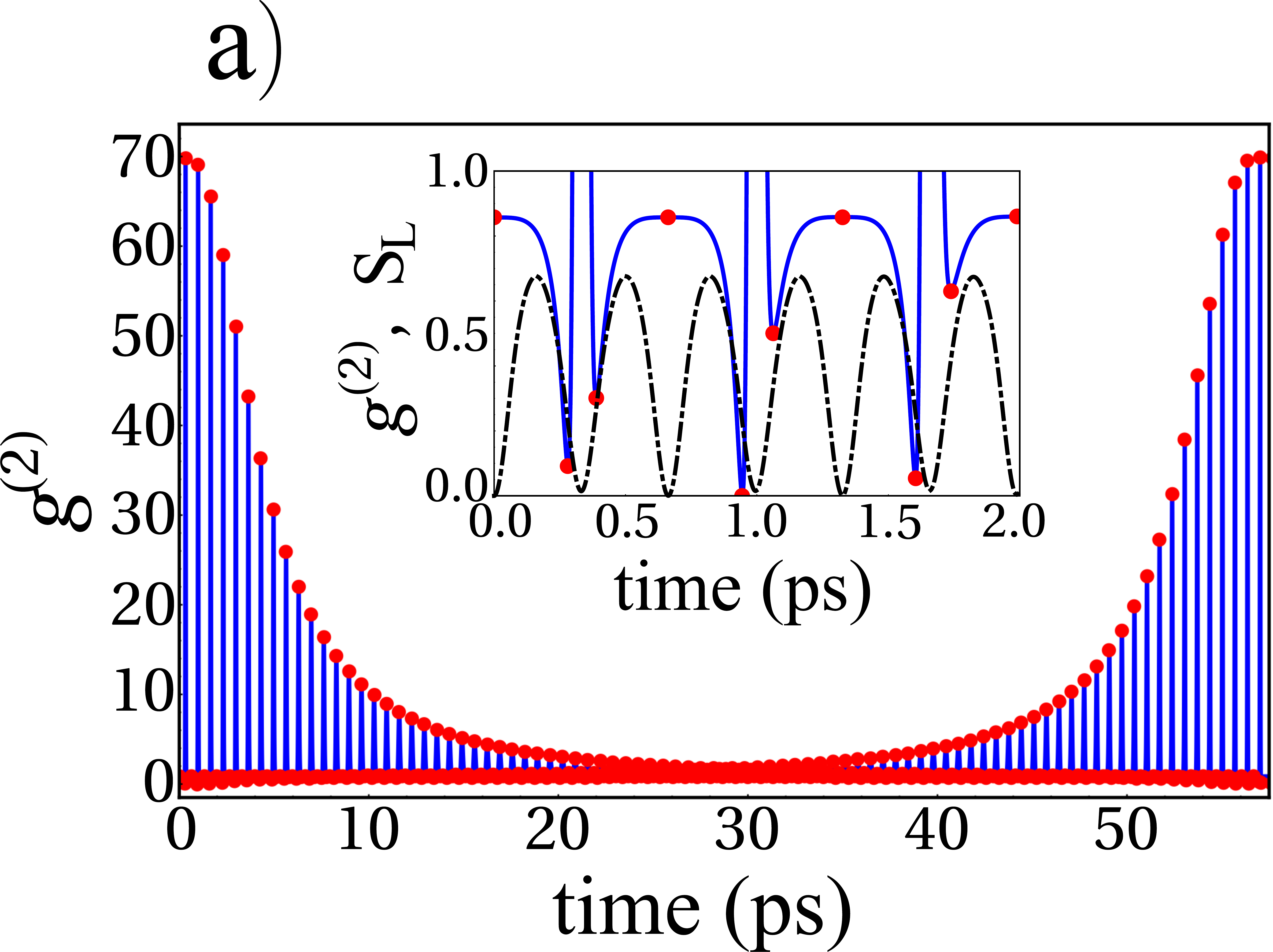}
\vspace{13pt}
\end{minipage}\\
\begin{minipage}[c]{\textwidth}
\centering
\includegraphics[width=0.54\textwidth]{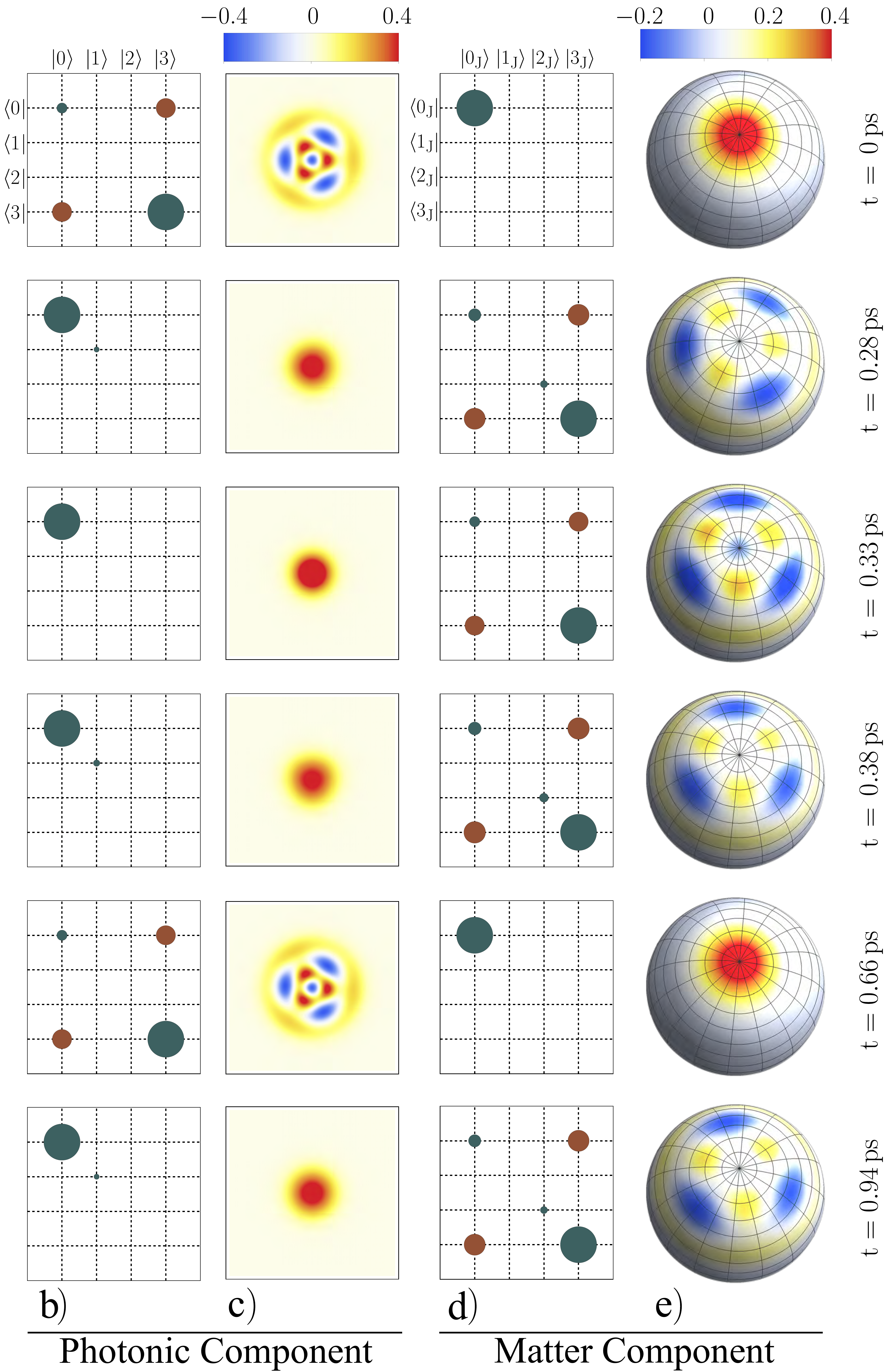}
\end{minipage}
\caption{System dynamics for $\mathrm{N}=10$ emitters and initial condition $\ket{\psi (0)}=\ket{\mathrm{M}}\otimes\ket{\bigstar}$, with $\mathrm{M}=-\mathrm{J}=-5$, $\eta=\sqrt{2}/3$ and $\beta=\sqrt{7}/3$. a) evolution of the photonic second-order correlation function $\mathrm{g}^{(2)}$ (blue solid line). Furthermore, the inset shows the linear entropy $\mathrm{S}_{\mathrm{L}}$ (black dot-dashed line) up to $\mathrm{t}=2$ ps. Red points depict the $\mathrm{g}^{(2)}$ critical values. b) matrix elements for the photonic density operator $\rho_{\mathrm{ph}}$. c) photonic Wigner functions, with $\mathrm{Re}(\epsilon)$ in the abscissa and $\mathrm{Im}(\epsilon)$ in the ordinate, where $\epsilon$ is a point in the phase space. Both abscissa and ordinate axes take values from $-3$ to $3$. d) matrix elements for the matter density operator $\rho_{\mathrm{D}}$. Here, was considered the notation $\ket{\mathrm{i}_{\mathrm{J}}}=\ket{\mathrm{-J+i}}$, with $\mathrm{i}=0,1,2,3$, for the matter states. e) matter Wigner distribution onto the block sphere. In panels b)-e) were considered the times of the first six critical values in the function $\mathrm{g}^{(2)}$.}
\label{graph3}
\end{figure}

To characterise the system we study the quantum evolution of experimental magnitudes like the second-order correlation function $\mathrm{g}^{(2)}$, the linear entropy $\mathrm{S}_{\mathrm{L}}$, and the Wigner function $\mathrm{W}$, in both cases for photons and matter. Figs. \ref{graph1}-\ref{graph3} show the dynamical process which transfer the photonic threefold star state into the equivalent \textit{Dicke threefold star} state. In Fig. \ref{graph1} we consider $\mathrm{N}=3$ emitters, and the initial condition is $\ket{\psi(\mathrm{t}=0)}=\eta\ket{\mathrm{M},0} + \beta \ket{\mathrm{M},3}$, with $\mathrm{J}=3/2$, $\mathrm{M}=-3/2$, $\eta=\sqrt{2}/3$ and $\beta=\sqrt{7}/3$. Fig \ref{graph1}.a) shows the evolution of the second-order correlation function (solid line) and the linear entropy (dot-dashed line) for the photonic reduced density matrix ($\rho_{\mathrm{ph}}$). The behaviour of both quantum experimental magnitudes $\mathrm{g}^{(2)}$ and $\mathrm{S}_{\mathrm{L}}$ repeats periodically, so we plotted this quantities up to their period $\mathrm{T}=6.32$ ps. We can observe that the second-order correlation function in this case takes values cyclically from the quantum regimen ($0\leqslant\mathrm{g}^{(2)}<1$) to superbunching regimen ($\mathrm{g}^{(2)}\gg1$), which is depicted by the red points.  \\

%%%

\begin{figure}[H]
\centering
\includegraphics[width=0.5\textwidth]{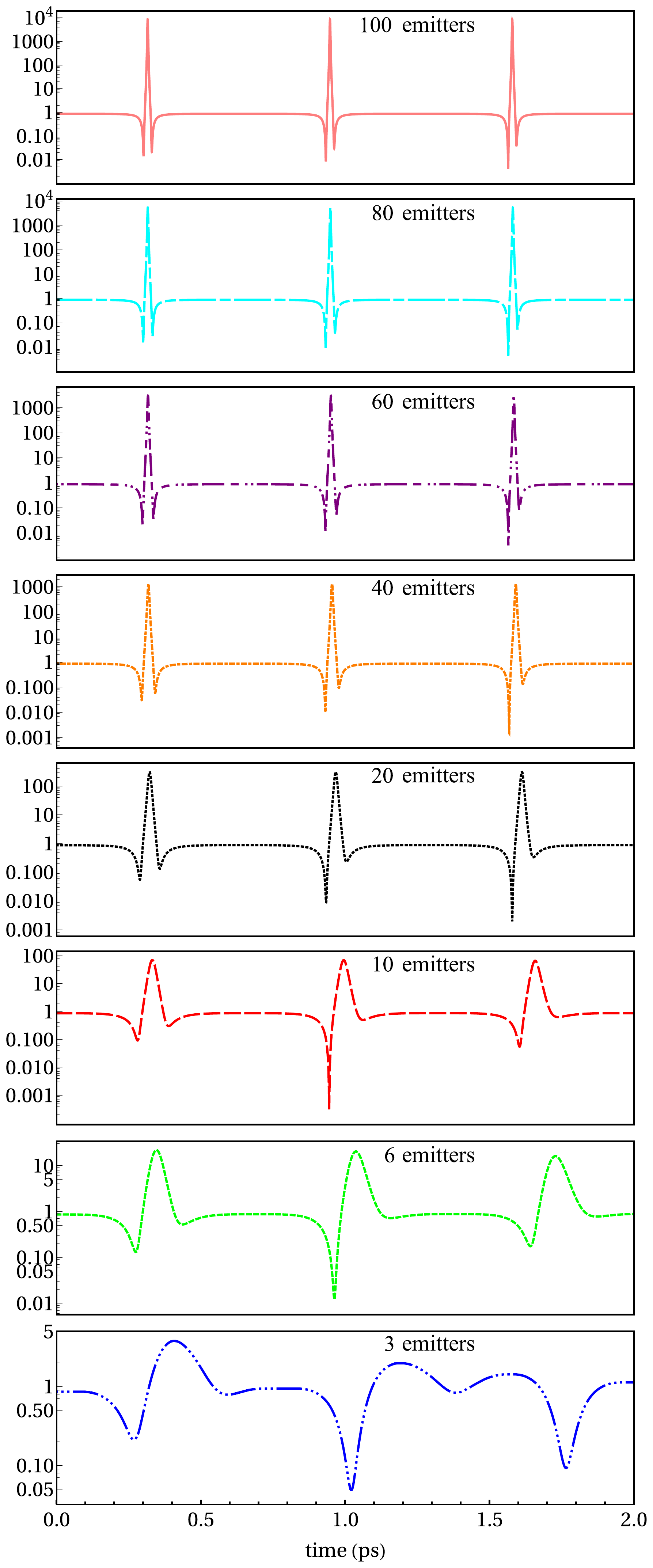}
\caption{Second-order correlation function $\mathrm{g}^{(2)}$ in logarithm scale, plotted up to $\mathrm{t}=2$ ps in order to describe the behaviour of the superbunching and the independence of the function $\mathrm{g}^{(2)}$ as the number of emitters increase. Panels are ordered from bottom to top for different number of emitters as $\mathrm{N}=3,6,10,20,40,60,80,100$.}
\label{graph4}
\end{figure}
%%%%%%
\newpage
\noindent

The linear entropy shows that the system never decouples completely. This fact entails that the state transfer is not efficient enough. Indeed, we can observe in the panel d) of the figure that there is a relevant component of the state $\ket{1_{\mathrm{J}}}=\ket{\mathrm{-J+1}}=\ket{-1/2}$ which never vanishes. On the other hand, the Wigner distribution becomes extremely delocalised over the Bloch sphere, making the threefold star state blurred, as can be seen in panel e).

In Fig. \ref{graph2} it is considered the case of $\mathrm{N}=6$ emitters, taking the same photonic threefold star state as initial condition (same $\eta$ and $\beta$ of the case $\mathrm{N}=3$), and $\mathrm{J}=3$, $\mathrm{M}=-3$ for the Dicke state. In Fig. \ref{graph2}.a), same as Fig \ref{graph1}.a), it is plotted the evolution of the second-order correlation function (up to its period $\mathrm{T}=18.35$ ps), which presents an increasing of its maximal values that means photons are reaching a stronger superbunching regimen. Furthermore, the inset shows, comparatively, the evolution of the linear entropy (dot-dashed line) up to $\mathrm{t}=2$ ps, and it is evidenced that the light-matter coupling is reduced around $\mathrm{t}=0.34$ ps and $\mathrm{t}=0.69$ ps, which is in agreement with the better transfer of the photonic threefold star state into the matter component of the system, as is also demonstrated in the more localised Wigner function in panel e), and the probability distributions in panel d), where the component of the state $\ket{1_{\mathrm{J}}}=\ket{-2}$ almost vanishes. Note that even though the threefold star state transfer to the matter from the photonic component is more efficient in the case $\mathrm{N}=6$, the minimal value of the linear entropy in $\mathrm{t}=0.34$ ps shows that light and matter are not completely decoupled.

With the purpose of study the system in a configuration such that guarantee the complete transfer of the threefold star state from the light to the matter component, we take into account an additional case of $\mathrm{N}=10$ emitters, considering the same photonic threefold star state (same $\eta$ and $\beta$) and  taking $\mathrm{J}=5$, $\mathrm{M}=-5$ for the initial Dicke matter state. From Fig. \ref{graph3}.a) it is possible to see that the light component reaches an increasingly strong superbunching regimen in comparison with the previous cases of $\mathrm{N}=3$ and $\mathrm{N}=6$ emitters; in addition, the linear entropy becomes periodical and its minimal values, which show the complete decoupling of light and matter, match with the optimal transfer of the threefold star state between the light and matter components, as can be seen in the inset (dot-dashed line) of the figure, and also in the well localised Wigner functions in 
%Fig. \ref{graph3} 
panels c) and e). This fact is reinforced by the probability distributions in panel d), where it can be noticed the threefold star state of matter completely transferred from the photonic counterpart, with only the states $\ket{0_{\mathrm{J}}}=\ket{\mathrm{-J}}$ and $\ket{3_{\mathrm{J}}}=\ket{\mathrm{-J+3}}$ contributing to the probability distribution in $\mathrm{t}=0.33$ ps.

Finally, the second-order correlation function is depicted, with time in the abscissa, comparatively in Fig~\ref{graph4}. Each panel is plotted for a fixed number of emitters $\mathrm{N}=3,6,10,20,40,60,80,100$. As is observed in the figure, the correlation function shape becomes independent of the number of emitters despite its value increases with $\mathrm{N}$. To understand this behaviour in a deeper way  we diagonalise the Hamiltonian matrix in equation~(\ref{eq:hamatrixlim}), which corresponds to the Hamiltonian subspace in the limit of $\mathrm{N}\rightarrow\infty$, yelding  $\lambda_1=-3\mathrm{g} \sqrt{2\mathrm{J}}-\Delta  \mathrm{J}$, $\lambda_2=- \mathrm{g} \sqrt{2\mathrm{J}}-\Delta  \mathrm{J}$, $\lambda_3= \mathrm{g}
   \sqrt{2\mathrm{J}}-\Delta  \mathrm{J}$, $\lambda_4=3  \mathrm{g} \sqrt{2\mathrm{J}}-\Delta  \mathrm{J}$ with eigenvectors given by the rows of the following matrix
\begin{equation}\label{eq:evecL}
\left(
\begin{array}{cccc}
 -1 & \sqrt{3} & -\sqrt{3} & 1 \\
 1 & -\frac{1}{\sqrt{3}} & -\frac{1}{\sqrt{3}} & 1 \\
 -1 & -\frac{1}{\sqrt{3}} & \frac{1}{\sqrt{3}} & 1 \\
 1 & \sqrt{3} & \sqrt{3} & 1 \\
\end{array}
\right).
\end{equation}

It is worth to notice that in this limit the eigenvectors of the Hamiltonian become independent of the angular momentum quantum number $\mathrm{J}$, i.e., they do not depend on the number of emitters in the assembly.  This fact allows us to conclude that all the observables become independent of the number of emitters. The expectation values $\braket{O}(\mathrm{t})$ evolves, from the operators in the Heisenberg picture $\dot{O}(\mathrm{t})=i[H,O]$, ruled by the Hamiltonian eigenstates. Therefore, the shape of the mean values of the observables depends on the parameters inherited by the Hamiltonian's eigenstates. 

\section{Conclusions}

The analysis of quantum dynamics of a system composed by an assembly of emitters embedded into a cavity is performed. We characterise the efficient transference of a threefold star light quantum state into the assembly.  As a general conclusion, we demonstrate that it is possible to transfer the photon state efficiently into the matter only in the limit of a  large number of emitters. The dressed states of photon and matter become independent of the number of two-level atoms yielding a shape invariant second-order correlation function where regimes of antibunching and superbunching appear cyclically.

\section*{Acknowledgements}

We would like to thank Boris A. Rodr\'iguez Rey for useful discussions. We acknowledge financial support from UN--DIEB project ``Control dinámico de la emisión en sistemas de Qubits acoplados con cavidades no-estaciona\-rias" HERMES 41611; and from COLCIENCIAS under the project ``Emi\-sión en sistemas de Qubits Superconductores acoplados a la radiación", Código 110171249692, CT 293-2016, HERMES 31361.  J.P.R.C. gratefully acknowledges financial support from the ``Beca de Doctorados Nacionales de COLCIENCIAS 785".

\bibliographystyle{elsarticle-num}
\bibliography{Ref.bib}

\end{document}